\documentclass[conference,10pt]{IEEEtran}
\IEEEoverridecommandlockouts
\usepackage{cite}
\usepackage{amsmath,amssymb,amsfonts}
\usepackage{algorithmic}
\usepackage{graphicx}
\usepackage{textcomp}
\usepackage{etoolbox}
\usepackage{float}

\floatstyle{ruled}
\newfloat{algorithm}{tbp}{loa}
\providecommand{\algorithmname}{Algorithm}
\floatname{algorithm}{\protect\algorithmname}

\def\BibTeX{{\rm B\kern-.05em{\sc i\kern-.025em b}\kern-.08em
    T\kern-.1667em\lower.7ex\hbox{E}\kern-.125emX}}

    \makeatletter
\patchcmd{\@maketitle}
  {\addvspace{0.5\baselineskip}\egroup}
  {\addvspace{-1\baselineskip}\egroup}
  {}
  {}

\makeatother

\begin{document}

\title{Optimizing Over-the-Air Computation in \\ IRS-Aided C-RAN Systems}

\author{\IEEEauthorblockN{Daesung Yu, Seok-Hwan Park} \IEEEauthorblockA{\textit{Dept. of Elect. Engineering} \\
 \textit{Jeonbuk National University}\\
 Jeonju, Korea \\
 \{imcreative93, seokhwan\}@jbnu.ac.kr} \and \IEEEauthorblockN{Osvaldo Simeone} \IEEEauthorblockA{\textit{KCLIP Lab, Centre for Telecomm. Research} \\
 \textit{Dept. of Engineering} \\
 \textit{King's College London}\\
 London, UK \\
 osvaldo.simeone@kcl.ac.uk} \and \IEEEauthorblockN{Shlomo Shamai (Shitz)} \IEEEauthorblockA{\textit{Dept. of Elect. Engineering} \\
 \textit{Technion}\\
 Haifa, Israel \\
sshlomo@ee.technion.ac.il} }
\maketitle
\begin{abstract}
Over-the-air computation (AirComp) is an efficient solution to enable federated learning on wireless channels. AirComp assumes that the wireless channels from different devices can be controlled, e.g., via transmitter-side phase compensation, in order to ensure coherent on-air combining. Intelligent reflecting surfaces (IRSs) can provide an alternative, or additional, means of controlling channel propagation conditions.
This work studies the advantages of deploying IRSs for AirComp systems in a large-scale cloud radio access network (C-RAN). In this system, worker devices upload locally updated models to a parameter server (PS) through distributed access points (APs) that communicate with the PS on finite-capacity fronthaul links. The problem of jointly optimizing the IRSs' reflecting phases and a linear detector at the PS is tackled with the goal of minimizing the mean squared error (MSE) of a parameter estimated at the PS.  Numerical results validate the advantages of deploying IRSs with optimized phases for AirComp in C-RAN systems.
\end{abstract}


\section{Introduction\label{sec:intro}}

\let\thefootnote\relax\footnotetext{This work was supported by Basic Science Research Program through the National Research Foundation of Korea grants funded by the Ministry of Education [NRF-2018R1D1A1B07040322, NRF-2019R1A6A1A09031717]. The work of O. Simeone was supported by the European Research Council (ERC) under the European Union's Horizon 2020 research and innovation programme (grant agreement No 725731). The work of S. Shamai was supported by the ERC under the European Union's Horizon 2020 research and innovation programme (grant agreement No 694630).}

Federated learning is an emerging distributed learning paradigm in which mobile devices collaboratively train a machine learning model while preserving the privacy of local data sets \cite{Bonawitz-et-al:arxiv19}. In the presence of latency and bandwidth constraints, the implementation of federated learning on wireless systems is challenging if many workers, or devices, are involved.
A potential solution to this problem is over-the-air computation (AirComp), which leverages the superposition property of the multiple access channel (MAC) from worker devices to a parameter server (PS) to allow for simultaneous transmissions from multiple devices \cite{Nazer-Gastpar:TIT07, Amiri-Gunduz:SPAWC19, Yang-et-al:TWC20}.
It was reported in \cite{Amiri-Gunduz:SPAWC19} that AirComp  outperforms a conventional multiple access technique in terms of test accuracy, and that the gain is particularly significant at low transmit power and large number of workers.

AirComp assumes that the wireless channels from different devices can be controlled, e.g., via transmitter-side phase compensation, in order to ensure coherent on-air combining \cite{Sery-Cohen:arxiv19}.
To alleviate this problem, the work \cite{Jiang-Shi:GC19} considered a deployment of  intelligent reflecting surfaces (IRSs).
IRSs, also referred to as reconfigurable intelligent surfaces, can be controlled through integrated electronics in order to shape their response to impinging electromagnetic waves \cite{Basar-et-al:Access19}. This enables the modification of the propagation channel between nearby transceivers. As a result, IRSs are considered as a cost-effective solution to improve spectral and energy efficiency of wireless systems \cite{Bjornson-et-al:arxiv19, Wu-Zhang:TWC19, Pan-et-al:arxiv19, DRenzo-et-al:EURASIP19}.
As examples of recent works on IRSs, references \cite{Wu-Zhang:TWC19} and \cite{Pan-et-al:arxiv19} addressed the joint design of downlink beamforming and IRSs' phases for interference management in multi-user \cite{Wu-Zhang:TWC19} and multi-cell systems \cite{Pan-et-al:arxiv19}. Reference \cite{Bjornson-et-al:arxiv19} analyzed the number of reflecting elements of IRSs needed to beat conventional wireless relaying techniques (see also \cite{Ntontin-et-al:arxiv19}).  Finally, an information-theoretic study was provided in \cite{Karasik-et-al:arxiv19}.

\begin{figure}
\centering
\!\!\!\!\!\!\!\!\centerline{\includegraphics[width=8.5cm, height=7.8cm, keepaspectratio]{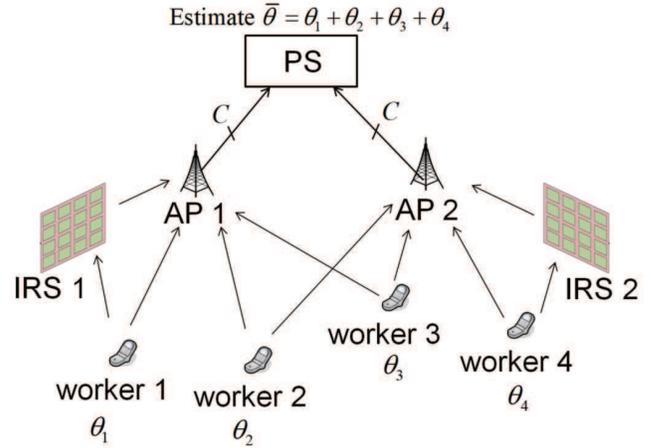}}
\caption{Over-the-air computation system in an IRS-aided C-RAN with $N_W=4$, $N_A=2$ and $N_I=2$.}
\label{fig:system-model}
\end{figure}

In this work, we study the advantages of deploying IRSs for AirComp systems. Unlike \cite{Jiang-Shi:GC19}, which focused on a MAC channel where workers directly communicate with a PS, we consider the large-scale cloud radio access network (C-RAN) illustrated in Fig. \ref{fig:system-model}, in which the workers upload local models to the PS through distributed access points (APs). The APs, or remote radio heads (RRHs), in C-RAN send the received signals to the PS on fronthaul links.
The fronthaul links have finite capacity, requiring fronthaul quantization and compression \cite{Park-et-al:SPM14}.
We tackle the problem of jointly optimizing the IRSs' reflecting phases and a linear detector at the PS with the goal of minimizing the mean squared error (MSE) of a parameter estimated at the PS. Due to the non-convexity of the problem, we propose an iterative algorithm that alternately updates the IRSs' phases and the linear detector. Via numerical results, we validate the advantages of deploying IRSs with optimized phases for AirComp in C-RAN systems.

\section{System Model}\label{sec:system}

As illustrated in Fig. \ref{fig:system-model}, we consider an over-the-air computation task performed on a C-RAN system. In the system, $N_W$ single-antenna worker devices send locally updated models to a PS through $N_A$ single-antenna APs. Each AP is connected to the PS via a fronthaul link, which we model as a digital link of capacity $C$ bit/sample \cite{Park-et-al:SPM14}. We define the sets $\mathcal{N}_W = \{1,2,\ldots,N_W\}$ and $\mathcal{N}_A = \{1,2,\ldots,N_A\}$ for the workers' and APs' indices, respectively.

\subsection{Over-the-Air Computation Model} \label{sub:over-the-air-computation}
We focus on the transmission at a specific time slot where each worker $k\in\mathcal{N}_W$ sends a scalar parameter  $\theta_k$, and the PS estimates a function $f(\boldsymbol{\theta})$ of the transmitted parameters $\boldsymbol{\theta}=\{\theta_k\}_{k\in\mathcal{N}_W}$. The parameter $\theta_k$ can be an element of the gradient vector \cite{Amiri-Gunduz:SPAWC19} or the local model \cite{Yang-et-al:TWC20} updated at worker $k$ using its local dataset. The PS typically estimates the weighted sum $f(\boldsymbol{\theta}) = \sum_{k\in\mathcal{N}_W} w_k \theta_k$, with $w_k = S_k/(\sum_{l\in\mathcal{N}_W} S_l)$, where $S_k$ denotes the number of training samples at device $k$ \cite{Yang-et-al:TWC20}. To simplify the discussion, we assume $S_k=S$ for all $k\in\mathcal{N}_W$, and that the target parameter denoted by $\bar{\theta}$ is given by the sum
\begin{align}
 \bar{\theta} = f(\boldsymbol{\theta})
 = \sum\nolimits_{k\in\mathcal{N}_W} \theta_k. \label{eq:target}
 \end{align}
We also assume that the parameters $\theta_k$ are independent, and we define the power of parameter $\theta_k$ as $\mathtt{E}[|\theta_k|^2]=\sigma_{\theta,k}^2$. Thus, the target parameter $\bar{\theta}$ has power $\mathtt{E}[|\bar{\theta}|^2] = \sum_{k\in\mathcal{N}_W} \sigma_{\theta,k}^2$.

\subsection{Channel Model} \label{sub:channel-model}
To assist edge communication from the workers to the APs, we assume the presence of $N_I$ IRSs \cite{Jiang-Shi:GC19} in the network. Each IRS has $n_I$ reflecting elements, whose reflecting phases are dynamically adjusted to adapt to the instantaneous channel state information (CSI). We define the set $\mathcal{N}_I=\{1,2,\ldots,N_I\}$ for the IRSs' indices.

Under a flat-fading channel model,  the received signal $y_i$ of AP $i$ can be written as
\begin{align}
    y_i = \sum\nolimits_{k\in\mathcal{N}_W} h_{i,k} x_k + z_i, \label{eq:received-signal-AP-i}
\end{align}
where $x_k$ is the signal transmitted by worker $k$; $h_{i,k}$ denotes the channel coefficient from worker $k$ to AP $i$; and $z_i\sim \mathcal{CN}(0,\sigma_z^2)$ represents the additive noise. The  signal $x_k$ satisfies the transmit power constraint $\mathtt{E}[|x_k|^2]\leq P$.

Due to the presence of IRSs, the channel coefficient $h_{i,k}$ is modelled as \cite{Bjornson-et-al:arxiv19,Wu-Zhang:TWC19,Pan-et-al:arxiv19}
\begin{align}
    h_{i,k} = \sqrt{\rho_{d,i,k}} h_{d,i,k} + \sum\nolimits_{j\in\mathcal{N}_I} \sqrt{\rho_{r,i,j,k}} \mathbf{g}_{i,j}^H \boldsymbol{\Theta}_j \mathbf{h}_{r,j,k}, \label{eq:composed-channel}
\end{align}
where $h_{d,i,k}$ denotes the small-scale fading channel  from worker $k$ to AP $i$; $\mathbf{g}_{i,j} \in \mathbb{C}^{n_I\times 1}$ represents the small-scale fading channel vector from IRS $j$ to AP $i$; $\mathbf{h}_{r,j,k} \in \mathbb{C}^{n_I\times 1}$ is the small-scale fading channel vector from worker $k$ to IRS $j$; $\rho_{d,i,k}$ denotes the path-loss of the direct link from worker $k$ to AP $i$; $\rho_{r,i,j,k}$ is the path-loss of the composite link from worker $k$ to AP $i$ through IRS $j$; and  $\mathbf{\Theta}_j$ is a diagonal matrix that represents the reflecting operation of IRS $j$, which is defined as\begin{align}
    \mathbf{\Theta}_j = \text{diag}\left(\{e^{j\phi_{j,m}}\}_{m=1}^{n_I}\right), \label{eq:reflecting-diagona-matrix}
\end{align}
where $\phi_{j,m}\in[0,2\pi)$ denotes the reflecting phase of the $m$th element of IRS $j$.

We model the path-loss $\rho_{d,i,k}$ between worker $k$ and AP $i$ as $\rho_{d,i,k} = c_0 \cdot \mathtt{D}( \mathbf{p}_{W,k}, \mathbf{p}_{A,i} )^{-\eta}$, where $\mathtt{D}(\mathbf{a},\mathbf{b}) = || \mathbf{a} - \mathbf{b} ||_2$ is the Euclidean distance in meter between the two input vectors, $\mathbf{p}_{W,k}$ and $\mathbf{p}_{A,i}$ denote the position vectors of worker $k$ and AP $i$, respectively, $\eta$ is the path-loss exponent, and $c_0$ denotes the path-loss at the reference distance of $1$ m.
For the path-loss $\rho_{r,i,j,k}$ of the composite channel from worker $k$ to AP $i$ through IRS $j$, we adopt the sum-distance model  \cite{Basar-et-al:Access19} which models $\rho_{r,i,j,k}$ as\begin{align}
    \rho_{r,i,j,k} =
    c_0\left( \mathtt{D}(\mathbf{p}_{W,k},\mathbf{p}_{I,j}) + \mathtt{D}(\mathbf{p}_{I,j},\mathbf{p}_{A,i}) \right)^{-\eta}, \label{eq:path-loss-composite}
\end{align}
where $\mathbf{p}_{I,j}$ denotes the position vector of IRS $j$.


\section{Over-the-Air Computation in IRS-aided C-RAN}

In this section, we illustrate the operations at the worker devices, the APs, and the PS in the IRS-aided C-RAN system described in Sec. \ref{sec:system}.

\subsection{Transmission at Worker Devices} \label{sub:transmission-workers}

Without claim of optimality (see \cite{Cao-et-al:arxiv19}), we assume that each worker $k$ uses the maximum transmit power $P$, so that the transmit signal $x_k$ is given as
\begin{align}
    x_k = \alpha_k \theta_k, \label{eq:transmit-power-control}
\end{align}
with the coefficient $\alpha_k = (P / \sigma_{\theta,k}^2)^{1/2}$.
We note that this does not require CSI at worker devices.

\subsection{Quantization at APs} \label{sub:quantization-APs}

AP $i$ sends a quantized version of the received signal $y_i$ to the PS through a fronthaul link of capacity $C$ bit/sample.
Under the assumptions that the updated model vectors have a sufficiently large dimension, the quantized signal denoted by $\hat{y}_i$ can be modelled as \cite{Park-et-al:SPM14,Zamir-Feder:TIT96}
\begin{align}
    \hat{y}_i = y_i + q_i, \label{eq:quantization-AP-i}
\end{align}
where $q_i$ models the quantization distortion as being independent of $y_i$ and distributed as $q_i\sim\mathcal{CN}(0,\omega_i)$.
According to standard rate-distortion theoretic results \cite{ElGamal-Kim:11}, the quantization noise power $\omega_i$ satisfies the condition
\begin{align}
    I(y_i ; \hat{y}_i) = \log_2\left( 1 + \frac{\sigma_{y,i}^2}{\omega_i} \right) \leq C, \label{eq:fronthaul-capacity-constraint}
\end{align}
where $\sigma_{y,i}^2$ denotes the variance of the received signal $y_i$ given as
\begin{align}
    \sigma_{y,i}^2 = \sum\nolimits_{k\in\mathcal{N}_W} |h_{i,k}|^2 P + \sigma_z^2. \label{eq:variance-y-i}
\end{align}
The minimum distortion power $\omega_i$ that satisfies the condition (\ref{eq:fronthaul-capacity-constraint}) is given as
\begin{align}
    \omega_i = \sigma_{y,i}^2/ (2^C - 1). \label{eq:minimum-omega-i}
\end{align}
Note that the optimal distortion level (\ref{eq:minimum-omega-i}) is a function of the reflecting phases $\boldsymbol{\phi}=\{\phi_{j,m}\}_{j\in\mathcal{N}_I,m\in\{1,\ldots,n_I\} }$, since $\boldsymbol{\phi}$ affects the channel coefficients $h_{i,k}$ as seen in (\ref{eq:composed-channel}).

\subsection{Estimation at PS} \label{sub:estimation-PS}

Based on the received quantized signals $\{\hat{y}_i\}_{i\in\mathcal{N}_A}$, the PS estimates the target parameter $\bar{\theta}$ in (\ref{eq:target}). To elaborate, let us define a vector $\hat{\mathbf{y}} = [\hat{y}_1 \, \hat{y}_2 \cdots \hat{y}_{N_A}]^T$ which stacks the quantized signals. Then, the vector $\hat{\mathbf{y}}$ can be expressed as
\begin{align}
    \hat{\mathbf{y}} = \sum\nolimits_{k\in\mathcal{N}_W} \alpha_k \mathbf{h}_k \theta_k + \mathbf{z} + \mathbf{q}, \label{eq:stacked-quantized-signal}
\end{align}
where we have defined the vectors $\mathbf{h}_k = [h_{1,k} \, h_{2,k} \cdots h_{N_A,k}]^T$, $\mathbf{z} = [z_1 \, z_2 \cdots z_{N_A}]^T \sim \mathcal{CN}(\mathbf{0}, \sigma_z^2\mathbf{I})$ and $\mathbf{q} = [q_1\, q_2 \cdots q_{N_A}]^T \sim \mathcal{CN}(\mathbf{0}, \mathbf{\Omega})$ with $\mathbf{\Omega} = \text{diag}(\{\omega_i\}_{i\in\mathcal{N}_A})$.

The channel vector $\mathbf{h}_k\in\mathbb{C}^{N_A \times 1}$ from worker $k$ to all the APs can be written as a function of the IRSs' phases $\boldsymbol{\phi}$ as
\begin{align}
    \mathbf{h}_k = \mathbf{h}_{d,k} + \sum\nolimits_{j\in\mathcal{N}_I} \mathbf{R}_{r,j,k} \mathbf{G}_j \text{diag}(\mathbf{h}_{r,j,k}) \mathbf{v}_j, \label{eq:detailed-hk}
\end{align}
where the matrices   $\mathbf{R}_{r,j,k}\in\mathbb{C}^{N_A\times N_A}$, $\mathbf{G}_j \in \mathbb{C}^{N_A\times n_I}$, and the vectors $\mathbf{h}_{d,k}\in\mathbb{C}^{N_A\times 1}$, $\mathbf{v}_j\in\mathbb{C}^{n_I\times 1}$ are defined as $\mathbf{R}_{r,j,k} = \text{diag}(\{\rho_{r,i,j,k}^{1/2}\}_{i\in\mathcal{N}_A})$, $\mathbf{G}_j = [\mathbf{g}_{1,j} \cdots \mathbf{g}_{N_A,j}]^H$, $\mathbf{h}_{d,k} = [\rho_{d,1,k}^{1/2}h_{d,1,k} \cdots \rho_{d,N_A,k}^{1/2}h_{d,
N_A,k}]^T$, and $\mathbf{v}_j = [e^{j\phi_{j,1}} \cdots e^{j\phi_{j,n_I}}]^T$, respectively.
Note that the optimization of the phases $\{\phi_{j,m}\}_{m=1}^{n_I}$ of IRS $j$ is equivalent to that of the vector $\mathbf{v}_j$ as long as the conditions
\begin{align}
    |\mathbf{v}_j(m)|^2 = 1 \label{eq:unit-modulus-constraint}
\end{align}
are satisfied for all $m\in\{1,\ldots,n_I\}$, where $\mathbf{v}_j(m)$ denotes the $m$th element of $\mathbf{v}_{j}$.
From the vector $\mathbf{v}_j$, each phase $\phi_{j,m}$ can be obtained as $-\angle\mathbf{v}_j(m)$.


We assume that the PS performs a linear estimation of the target parameter $\bar{\theta}$ from $\hat{\mathbf{y}}$. Accordingly, an estimate $\hat{\bar{\theta}}$ of $\bar{\theta}$ is given as
\begin{align}
    \hat{\bar{\theta}} = \mathbf{f}^H \hat{\mathbf{y}}, \label{eq:linear-estimate}
\end{align}
with a linear detection vector  $\mathbf{f}\in\mathbb{C}^{N_A\times 1}$.

For given phases $\boldsymbol{\phi}$, i.e., $\mathbf{v} = \{\mathbf{v}_j\}_{j\in\mathcal{N}_I}$, and linear detection vector $\mathbf{f}$, the MSE between the estimate $\hat{\bar{\theta}}$ and the target parameter $\bar{\theta}$ is evaluated as
\begin{align}
    \mathtt{e}(\mathbf{v}, \mathbf{f}) &= \mathtt{E}\left[ |\hat{\bar{\theta}} - \bar{\theta}|^2 \right] \label{eq:MSE} \\
     &= \sum_{k\in\mathcal{N}_W} | \alpha_k \mathbf{f}^H \mathbf{h}_k - 1 |^2\sigma_{\theta,k}^2 + \mathbf{f}^H \left( \sigma_z^2\mathbf{I} + \mathbf{\Omega} \right) \mathbf{f}. \nonumber
\end{align}

\section{Optimization} \label{sec:optimization}

We tackle the problem of jointly optimizing the IRSs' reflecting phases $\mathbf{v}$ and the linear detection vector $\mathbf{f}$ of the PS with the goal of minimizing the MSE $\mathtt{e}(\mathbf{v}, \mathbf{f})$ in (\ref{eq:MSE}) while satisfying the unit modulus constraints (\ref{eq:unit-modulus-constraint}).
The problem can be stated as
\begin{subequations} \label{eq:problem-original}
\begin{align}
    \underset{\mathbf{v},\mathbf{\mathbf{f}}}{\mathrm{minimize}} \,\,\, & \mathtt{e}(\mathbf{v}, \mathbf{f}) \label{eq:problem-original-cost} \\
    \text{s.t.}\,\,\,\,\,\, & |\mathbf{v}_j(m)|^2 = 1, \, j\in\mathcal{N}_I, \, m\in\{1,\ldots,n_I\}. \label{eq:problem-original-unit-modulus}
\end{align}
\end{subequations}
Since it is difficult to jointly optimize the variables $\mathbf{v}$ and $\mathbf{f}$, we propose an iterative algorithm that alternately optimizes one variable while fixing other.


If we fix the IRSs' phases $\mathbf{v}$ in problem (\ref{eq:problem-original}), finding the optimal detector $\mathbf{f}$ becomes an unconstrained quadratic optimization problem, whose closed-form solution is given as
\begin{align}
    \mathbf{f} = \left( \sum_{k\in\mathcal{N}_W}P\mathbf{h}_k\mathbf{h}_k^H + \sigma_z^2\mathbf{I} + \mathbf{\Omega} \right)^{-1} \sum_{k\in\mathcal{N}_W}\alpha_k \sigma_{\theta,k}^2\mathbf{h}_k \,. \label{eq:optimal-linear-detector}
\end{align}


To tackle the problem of optimizing the IRSs' phases $\mathbf{v}$ for fixed $\mathbf{f}$, we remove the terms that are not dependent on the IRSs' phases from the cost function. Stating the obtained problem with respect to a stacked vector
$\bar{\mathbf{v}} = [\mathbf{v}_1^H \mathbf{v}_2^H \cdots \mathbf{v}_{N_I}^H]^H \in\mathbb{C}^{\bar{n}_I\times 1}$ with $\bar{n}_I=n_I N_I$ yields
\begin{subequations} \label{eq:problem-vBar}
\begin{align}
    \underset{\bar{\mathbf{v}}}{\mathrm{minimize}} \,\,\, &\!\! \left(\!\!\! \begin{array}{c} \sum_{k\in\mathcal{N}_W} \! \left( |\mathbf{a}_k^H\bar{\mathbf{v}}|^2 \!+\! 2\Re\{b_k^* \mathbf{a}_k^H\bar{\mathbf{v}}\} \right)\sigma_{\theta,k}^2 + \\
    \sum_{i\in\mathcal{N}_A, k\in\mathcal{N}_W} \! \left( |\mathbf{c}_{i,k}^H\bar{\mathbf{v}}|^2 \!+\! 2\Re\{d_{i,k}^*\mathbf{c}_{i,k}^H\bar{\mathbf{v}}\} \right)
    \end{array} \!\!\! \right)
    \label{eq:problem-vBar-cost} \\
    \text{s.t.}\,\,\,\,\,\, & |\bar{\mathbf{v}}(m)|^2 = 1, \, m\in\{1,\ldots,\bar{n}_I\}, \label{eq:problem-vBar-modulus-constraint}
\end{align}
\end{subequations}
where we have defined the notations $\mathbf{a}_k = \bar{\mathbf{H}}_{r,k}^H\mathbf{f}\alpha_k^*\in\mathbb{C}^{\bar{n}_I\times 1}$, $b_k = \alpha_k \mathbf{f}^H\mathbf{h}_{d,k} - 1$, $\bar{\mathbf{H}}_{r,k} = [\mathbf{H}_{r,1,k} \mathbf{H}_{r,2,k} \cdots \mathbf{H}_{r,N_I,k}]\in\mathbb{C}^{N_A \times \bar{n}_I}$, $\mathbf{H}_{r,j,k} = \mathbf{R}_{r,j,k} \mathbf{G}_j \text{diag}(\mathbf{h}_{r,j,k}) \in\mathbb{C}^{N_A\times n_I}$, $\mathbf{c}_{i,k} = (P/(2^C-1))^{1/2} |\mathbf{f}(i)| \bar{\mathbf{H}}_{r,k}^H\mathbf{e}_i \in\mathbb{C}^{\bar{n}_I\times 1}$, and $d_{i,k} = (P/(2^C-1))^{1/2} |\mathbf{f}(i)|
 \mathbf{e}_i^H\mathbf{h}_{d,k}$ with $\mathbf{f}(i)$ and $\mathbf{e}_i$ being the $i$th element of $\mathbf{f}$ and the $i$th column of an identity matrix of size $N_A$, respectively.

The problem (\ref{eq:problem-vBar}) is non-convex due to the unit modulus constraints (\ref{eq:problem-vBar-modulus-constraint}). To handle this issue, we adopt the matrix lifting approach proposed in \cite{Jiang-Shi:GC19}. Accordingly, we tackle the problem (\ref{eq:problem-vBar}) with respect to a matrix $\mathbf{V} \in \mathbb{C}^{(\bar{n}_I+1) \times (\bar{n}_I+1)}$ defined as
\begin{align}
    \mathbf{V} = \left[\begin{array}{c} \bar{\mathbf{v}} \\ 1 \end{array}\right] \left[\bar{\mathbf{v}}^H \,\,1 \right] = \left[\begin{array}{cc} \bar{\mathbf{v}}\bar{\mathbf{v}}^H & \bar{\mathbf{v}} \\ \bar{\mathbf{v}}^H & 1 \end{array}\right]. \label{eq:quadratic-variable}
\end{align}
The matrix $\mathbf{V}$ is subject to the constraints $\mathbf{V}\succeq\mathbf{0}$, $\text{rank}(\mathbf{V}) \leq 1$, and $\mathbf{V}(m,m)=1$ for all $m\in\{1,2,\ldots,\bar{n}_I+1\}$. From $\mathbf{V}$, the IRSs' phase vector $\bar{\mathbf{v}}$ can be recovered as the first $\bar{n}_I$ elements of the last column of $\mathbf{V}$.

We tackle (\ref{eq:problem-vBar}) with respect to $\mathbf{V}$ by using the following equalities:
\begin{align}
    |\mathbf{a}_k^H\bar{\mathbf{v}}|^2 + 2\Re\{\mathbf{b}_k^* \mathbf{a}_k^H\bar{\mathbf{v}}\} & = \left[ \bar{\mathbf{v}}^H \,\, 1 \right]\left[ \begin{array}{cc} \mathbf{a}_k \mathbf{a}_k^H & b_k\mathbf{a}_k \\ b_k^*\mathbf{a}_k^H & 0 \end{array} \right] \left[ \begin{array}{c} \bar{\mathbf{v}} \\ 1 \end{array} \right] \nonumber \\
    & = \text{tr}\left( \left[ \begin{array}{cc} \mathbf{a}_k \mathbf{a}_k^H & b_k\mathbf{a}_k \\ b_k^*\mathbf{a}_k^H & 0 \end{array} \right] \mathbf{V} \right), \label{eq:cost-quadratic1} \\
    |\mathbf{c}_{i,k}^H\bar{\mathbf{v}}|^2 + 2\Re\{\mathbf{d}_{i,k}^* \mathbf{c}_{i,k}^H\bar{\mathbf{v}}\} & = \left[ \bar{\mathbf{v}}^H \,\, 1 \right]\!\!\left[\! \begin{array}{cc} \mathbf{c}_{i,k} \mathbf{c}_{i,k}^H & d_{i,k}\mathbf{c}_{i,k} \\ d_{i,k}^*\mathbf{c}_{i,k}^H & 0 \end{array} \!\right] \!\! \left[\! \begin{array}{c} \bar{\mathbf{v}} \\ 1 \end{array} \! \right] \nonumber \\
    & = \text{tr}\left( \left[ \begin{array}{cc} \mathbf{c}_{i,k} \mathbf{c}_{i,k}^H & d_{i,k}\mathbf{c}_{i,k} \\ d_{i,k}^*\mathbf{c}_{i,k}^H & 0 \end{array} \right] \mathbf{V} \right). \label{eq:cost-quadratic2}
\end{align}
Specifically, by substituting (\ref{eq:cost-quadratic1}) and (\ref{eq:cost-quadratic2}) into problem (\ref{eq:problem-vBar}), we obtain the problem\begin{subequations} \label{eq:problem-V}
\begin{align}
    \underset{\mathbf{V}\succeq \mathbf{0}}{\mathrm{minimize}} \,\,\, & \text{tr}\left( \mathbf{M}\mathbf{V} \right) \label{eq:problem-V-cost} \\
    \text{s.t.}\,\,\,\,\,\, & \mathbf{V}(m,m) = 1, \, m\in\{1,\ldots,\bar{n}_I+1\},  \label{eq:problem-V-modulus-constraint} \\
    & \text{rank}(\mathbf{V}) \leq 1, \label{eq:problem-V-rank-constraint}
\end{align}
\end{subequations}
with the matrix $\mathbf{M}$ defined as
\begin{align}
    \mathbf{M} & = \sum_{k\in\mathcal{N}_W} \sigma_{\theta,k}^2 \left[ \begin{array}{cc} \mathbf{a}_k \mathbf{a}_k^H & b_k\mathbf{a}_k \\ b_k^*\mathbf{a}_k^H & 0 \end{array} \right] \label{eq:definition-M} \\
    & + \sum_{i\in\mathcal{N}_A,k\in\mathcal{N}_W} \left[ \begin{array}{cc} \mathbf{c}_{i,k} \mathbf{c}_{i,k}^H & d_{i,k}\mathbf{c}_{i,k} \\ d_{i,k}^*\mathbf{c}_{i,k}^H & 0 \end{array} \right].  \nonumber
\end{align}

To address the non-convexity of constraint (\ref{eq:problem-V-rank-constraint}), we note that (\ref{eq:problem-V-rank-constraint}) is equivalent to the constraint \cite{Jiang-Shi:GC19}
\begin{align}
    \text{tr}(\mathbf{V}) - \sigma_1(\mathbf{V}) = 0, \label{eq:rank-1-constraint-rewritten}
\end{align}
where $\sigma_1(\cdot)$ denotes the largest singular value of the input matrix. Function $\sigma_1(\mathbf{V})$ is convex in $\mathbf{V}$ \cite{Boyd:04}. Furthermore, for $\mathbf{V}\succeq \mathbf{0}$, the left-hand side (LHS) of (\ref{eq:rank-1-constraint-rewritten}) is 0 when $\text{rank}(\mathbf{V})\leq 1$ and it becomes larger than 0 otherwise.

Based on this observation, as in \cite{Jiang-Shi:GC19}, we tackle the problem\begin{subequations} \label{eq:problem-penalty}
\begin{align}
    \underset{\mathbf{V}\succeq \mathbf{0}}{\mathrm{minimize}} \,\,\, & \text{tr}\left( \mathbf{M}\mathbf{V} \right) + \gamma\left( \text{tr}(\mathbf{V}) - \sigma_1(\mathbf{V}) \right) \label{eq:problem-penalty-cost} \\
    \text{s.t.}\,\,\,\,\,\, & \mathbf{V}(m,m) = 1, \, m\in\{1,\ldots,\bar{n}_I+1\},  \label{eq:problem-penalty-modulus-constraint}
\end{align}
\end{subequations}
with a fixed weight $\gamma\geq 0$.
In problem (\ref{eq:problem-penalty}), we have removed the rank constraint (\ref{eq:problem-V-rank-constraint}) and instead added a penalty term $\gamma( \text{tr}(\mathbf{V}) - \sigma_1(\mathbf{V}) )$ to the cost function that increases if (\ref{eq:problem-V-rank-constraint}) is not satisfied.

The problem (\ref{eq:problem-penalty}) is a difference-of-convex (DC) problem whose locally optimal solution can be efficiently found via the concave convex procedure (CCP) approach \cite{Tao-et-al:TWC16}.
CCP solves a sequence of convex problems obtained by linearizing the terms that induce non-convexity. In the DC problem (\ref{eq:problem-penalty}), the only term that induces non-convexity is $-\gamma\cdot \sigma_1(\mathbf{V})$ in the penalty term. Linearizing $-\gamma\cdot\sigma_1(\mathbf{V})$ at a reference point  $\mathbf{V} = \mathbf{V}^{\prime}$ yields the upper bound \cite{Jiang-Shi:GC19}
\begin{align}
    -\gamma\cdot\sigma_1(\mathbf{V}) \leq -\gamma\cdot\text{tr}\left( \mathbf{V}\, \mathbf{u}_1(\mathbf{V}^{\prime})\mathbf{u}_1(\mathbf{V}^{\prime})^H \right), \label{eq:linearizing-spectral-norm}
\end{align}
where $\mathbf{u}_1(\cdot)$ returns the eigenvector of the input matrix corresponding to the largest eigenvalue. The condition (\ref{eq:linearizing-spectral-norm}) is satisfied with equality when $\mathbf{V} = \mathbf{V}^{\prime}$. The CCP based algorithm for optimizing $\mathbf{V}$ is summarized in Algorithm 1.
\begin{algorithm}
\caption{CCP based algorithm for optimizing $\mathbf{V}$}

\textbf{1.} Initialize $\mathbf{V}^{(1)}$ as (\ref{eq:quadratic-variable}) with arbitrary $\bar{\mathbf{v}}$ that satisfies (\ref{eq:problem-vBar-modulus-constraint}), and set $t\leftarrow 1$

\textbf{2.} Update $\mathbf{V}^{(t+1)}$ as a solution of the convex problem:
\begin{align} 
    \underset{\mathbf{V}\succeq \mathbf{0}}{\mathrm{minimize}} \,\,\, & \text{tr}\left( \mathbf{M}\mathbf{V} \right) \!+\! \gamma\left(\! \text{tr}(\mathbf{V}) - \text{tr}\left( \mathbf{V}\, \mathbf{u}_1(\mathbf{V}^{(t)})\mathbf{u}_1(\mathbf{V}^{(t)})^H \right)\! \right) \nonumber \\ 
    \text{s.t.}\,\,\,\,\,\, & \mathbf{V}(m,m) = 1, \, m\in\{1,\ldots,\bar{n}_I+1\},  \nonumber 
\end{align}

\textbf{3.} Stop if $|| \mathbf{V}^{(t+1)} - \mathbf{V}^{(t)} ||_F^2 \leq \delta$ is satisfied. Otherwise, go back to Step 2 with $t\leftarrow t+1$.
\end{algorithm}


Overall, the proposed algorithm that alternately optimizes the IRSs' phases $\mathbf{v}$ and the linear detector $\mathbf{f}$ is detailed in Algorithm 2.
In the algorithm, we initialize $\mathbf{v}$ and $\mathbf{f}$ in Steps 1-2, and update $\mathbf{v}$ for fixed $\mathbf{f}$ in Steps 3-4. In Step 4, $\mathbf{v}$ is modified only when it does not satisfy the modulus constraints (\ref{eq:problem-vBar-modulus-constraint}). In Step 5, $\mathbf{f}$ is updated for fixed $\mathbf{v}$, and we check the convergence in Step 6.

\begin{algorithm}
\caption{Proposed algorithm alternately optimizing $\mathbf{v}$ and $\mathbf{f}$}

\textbf{1.} Initialize $\mathbf{v}^{(1)}$ as arbitrary vectors that satisfy (\ref{eq:problem-original-unit-modulus}).

\textbf{2.} Update $\mathbf{f}^{(1)}$ according to (\ref{eq:optimal-linear-detector}) with $\mathbf{v}=\mathbf{v}^{(1)}$, and set $t\leftarrow 1$.

\textbf{3.} For $j\in\mathcal{N}_I$, update $\mathbf{v}_j^{(t+1)}$ as the elements from $(j-1)n_I+1$ to $j n_I$ of the last column of the matrix $\mathbf{V}$ obtained by Algorithm 1 with $\mathbf{f} = \mathbf{f}^{(t)}$.

\textbf{4.} For $j\in\mathcal{N}_I$ and $m\in\{1,\ldots,n_I\}$, update $\mathbf{v}_j^{(t+1)}(m)\leftarrow \mathbf{v}_j^{(t+1)}(m)/|\mathbf{v}_j^{(t+1)}(m)|$.

\textbf{5.} Update $\mathbf{f}^{(t+1)}$ according to (\ref{eq:optimal-linear-detector}) with $\mathbf{v}=\mathbf{v}^{(t+1)}$.

\textbf{6.} Stop if $\sum_{j\in\mathcal{N}_I}|| \mathbf{v}_j^{(t+1)} - \mathbf{v}_j^{(t)} ||^2 + ||\mathbf{f}^{(t+1)} - \mathbf{f}^{(t)}||^2 \leq \delta$. Otherwise, go back to Step 3 with $t\leftarrow t+1$.

\end{algorithm}

\section{Numerical Results} \label{sec:numerical-results}

In simulation, we assume that the positions of $N_W$ workers, $N_A$ APs and $N_I$ IRSs are uniformly distributed in a circular area of radius 100 m. We set the variance of local parameters to $\sigma_{\theta,k}^2 = 1$ for $k\in\mathcal{N}_W$ and
assume $c_0 = 20$ dB, $\eta=3$ in the path-loss models and $\gamma=1$ for the penalty coefficient in (\ref{eq:problem-penalty-cost}).
For all links, we consider independent Rayleigh fading channels that are distributed as $h_{d,i,k}\sim \mathcal{CN}(0,1)$, $\mathbf{g}_{i,j}\sim\mathcal{CN}(\mathbf{0}, \mathbf{I})$ and $\mathbf{h}_{r,j,k}\sim\mathcal{CN}(\mathbf{0}, \mathbf{I})$. We compare the performance of the proposed optimized scheme with two baseline schemes, one without IRSs and one with IRSs whose reflecting phases are randomly chosen. In all figures, we plot the normalized MSE, which is defined as the MSE $\mathtt{e}(\mathbf{v}, \mathbf{f})$ normalized by $\mathtt{E}[|\bar{\theta}|^2]$ so that it lies in the range $[0,1]$.

\begin{figure}
\centering
\!\!\!\!\!\!\!\!\centerline{\includegraphics[width=8.8cm, height=7.55cm]{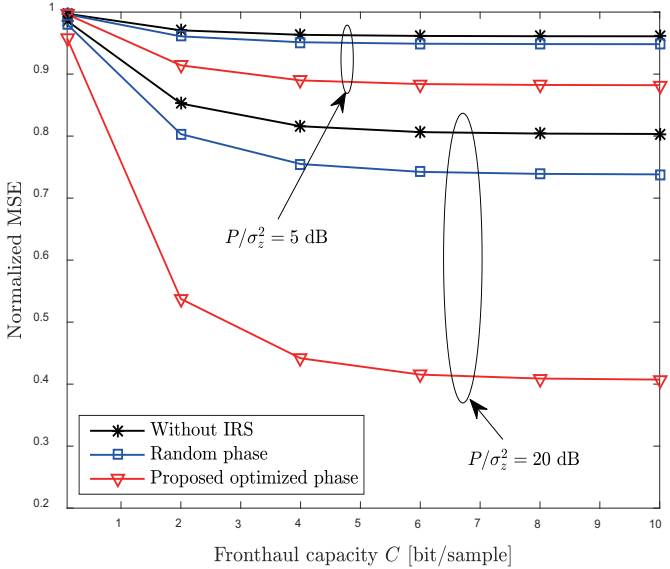}}
\caption{Average normalized MSE versus the fronthaul capacity $C$ for an IRS-aided C-RAN with $N_W=10$, $N_A=5$, $N_I=2$, $n_I=10$ and $P/\sigma_z^2\in\{5,20\}$ dB.}
\label{fig:graph-NMSE-vs-C}
\end{figure}

In Fig. \ref{fig:graph-NMSE-vs-C}, we plot the average normalized MSE versus the fronthaul capacity $C$ for an IRS-aided C-RAN system with $N_W=10$, $N_A=5$, $N_I=2$, $n_I=10$ and $P/\sigma_z^2\in\{5,20\}$ dB. The figure shows that the proposed optimized scheme outperforms both baseline schemes without IRS and with random phases, and that the gain increases with the fronthaul capacity $C$.
This is because, when $C$ is small, the impact of carefully designing the IRSs' phases becomes minor due to the impact of the quantization noise signals $\{q_i\}_{i\in\mathcal{N}_A}$.
Also, the gain increases with the signal-to-noise ratio (SNR) $P/\sigma_z^2$ of the uplink channel, and this trend coincides with the observation reported in \cite[Sec. IV]{Pan-et-al:arxiv19}.

\begin{figure}
\centering
\!\!\!\!\!\!\!\!\centerline{\includegraphics[width=8.8cm, height=7.55cm]{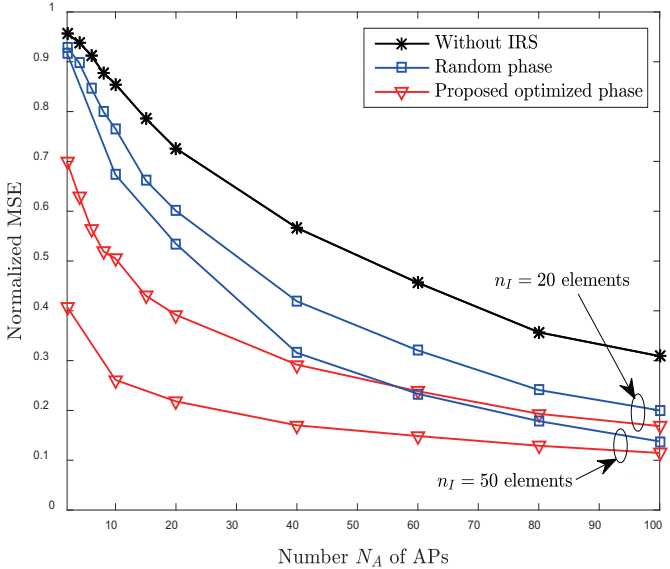}}
\caption{Average normalized MSE versus the number $N_A$ of APs for an IRS-aided C-RAN system with $N_W=5$, $N_I=2$, $n_I\in\{20,50\}$, $C=5$ and $P/\sigma_z^2=10$ dB.}
\label{fig:graph-NMSE-vs-NA}
\end{figure}

Fig. \ref{fig:graph-NMSE-vs-NA} plots the average normalized MSE versus the number $N_A$ of APs for an IRS-aided C-RAN system with $N_W=5$, $N_I=2$, $n_I\in\{20,50\}$, $C=5$ and $P/\sigma_z^2=10$ dB.
When there are only a few APs, deploying IRSs provides  relevant gains only when the reflecting phases are optimized according to Algorithm 2. However, the impact of optimizing the reflecting phases becomes minor for sufficiently large $N_A$.

\section{Concluding Remarks}\label{sec:conclusion}

We have studied the impacts of deploying IRSs on AirComp  in a C-RAN system.
To this end, we have tackled the joint optimization of the IRSs' reflecting phases and the linear detector at the PS with the goal of minimizing the MSE of the parameter estimated at the PS. Numerical results were provided that investigate the effects of various parameters on the performance gain of the proposed optimization scheme compared to baseline schemes.
Among open problems, we mention the design of channel estimation process, the investigation of the effect of imperfect CSI, and the design of AirComp jointly with information transfer.

\end{document}